\documentclass[12pt]{article}

\usepackage[dvipdfmx]{graphicx}
\usepackage{amsfonts}
\usepackage[mathscr]{eucal}
\usepackage{ascmac}
\usepackage{dcolumn}
\usepackage{bm}
\usepackage[colorlinks=true,linkcolor=blue,citecolor=blue]{hyperref}
\usepackage{color}
\usepackage{amsmath}

\begin{document}


\newcommand{\gtrsim}{ \mathop{}_{\textstyle \sim}^{\textstyle >} }
\newcommand{\lesssim}{ \mathop{}_{\textstyle \sim}^{\textstyle <} }
\newcommand{\vev}[1]{ \left\langle {#1} \right\rangle }
\newcommand{\bra}[1]{ \langle {#1} | }
\newcommand{\ket}[1]{ | {#1} \rangle }
\newcommand{\EV}{ \ {\rm eV} }
\newcommand{\KEV}{ \ {\rm keV} }
\newcommand{\MEV}{\  {\rm MeV} }
\newcommand{\GEV}{\  {\rm GeV} }
\newcommand{\TEV}{\  {\rm TeV} }
\newcommand{\1}{\mbox{1}\hspace{-0.25em}\mbox{l}}
\newcommand{\Red}[1]{{\color{red} {#1}}}

\newcommand{\lmk}{\left(}  
\newcommand{\rmk}{\right)}
\newcommand{\lkk}{\left[}  
\newcommand{\rkk}{\right]}
\newcommand{\lhk}{\left \{ }  
\newcommand{\rhk}{\right \} }
\newcommand{\del}{\partial}  
\newcommand{\la}{\left\langle} 
\newcommand{\ra}{\right\rangle}
\newcommand{\half}{\frac{1}{2}}

\newcommand{\bea}{\begin{array}}
\newcommand{\eea}{\end{array}}
\newcommand{\beq}{\begin{eqnarray}}
\newcommand{\eeq}{\end{eqnarray}}

\newcommand{\dd}{\mathrm{d}}
\newcommand{\Mpl}{M_{\rm Pl}}
\newcommand{\mg}{m_{3/2}}
\newcommand{\abs}[1]{\left\vert {#1} \right\vert}
\newcommand{\mphi}{m_{\phi}}
\newcommand{\Hz}{\ {\rm Hz}}
\newcommand{\for}{\quad \text{for }}
\newcommand{\Min}{\text{Min}}
\newcommand{\Max}{\text{Max}}
\newcommand{\Kahler}{K\"{a}hler }
\newcommand{\cphi}{\varphi}

\begin{titlepage}

\baselineskip 8mm

\begin{flushright}
IPMU 16-0049, TU-1020
\end{flushright}

\begin{center}

\vskip 1.2cm

{\Large\bf Charged Q-ball Dark Matter from $B$ and $L$ direction}

\vskip 1.8cm

{\large 
Jeong-Pyong Hong$^{a,b}$, 
Masahiro Kawasaki$^{a,b}$, 
and
Masaki Yamada$^{a,b,c}$}

\vskip 0.4cm

{\it$^a$Institute for Cosmic Ray Research, The University of Tokyo,
5-1-5 Kashiwanoha, Kashiwa, Chiba 277-8582, Japan}\\
{\it$^b$Kavli IPMU (WPI), UTIAS, The University of Tokyo, 5-1-5 Kashiwanoha, 
Kashiwa, 277-8583, Japan}\\
{\it$^c$Department of Physics, Tohoku University, Sendai, Miyagi 980-8578, Japan}

\date{\today}
\vspace{2cm}

\begin{abstract}  
We consider nearly equal number of gauge mediation type charged (anti-) Q-balls with charge of $\alpha^{-1}\simeq137$ well before the BBN epoch and discussed how they evolve in time. We found that ion-like objects with electric charges of $+O(1)$ are likely to become relics in the present universe, which we expect to be the dark matter. These are constrained by MICA experiment, where the trail of heavy atom-like or ion-like object in $10^9$ years old ancient mica crystals is not observed. We found that the allowed region for gauge mediation model parameter and reheating temperature have to be smaller than the case of the neutral Q-ball dark matter. 
\end{abstract}


\end{center}
\end{titlepage}

\baselineskip 6mm


\section{Introduction
\label{sec:introduction}}
Affleck-Dine mechanism~\cite{ad} is a promising candidate for baryogenesis based on supersymmetric (SUSY) theories due to its consistency with the observational bound on reheating temperature which avoids the gravitino problem~\cite{kkrehgrav}. In the Affleck-Dine mechanism, baryon number is generated from rotation in the phase direction of baryonic scalar field such as squark, which we call Affleck-Dine field, since baryon number has the meaning of angular momentum in complex plane of the field. After the Affleck-Dine mechanism, the spatial inhomogeneities of the Affleck-Dine field due to quantum fluctuations grows exponentially into non-topological solitons, which are called Q-balls~\cite{ks,kkins,kkins2}. They are defined as spherical solutions in a global $U(1)$ theory which minimize energy of the system with $U(1)$ charge fixed~\cite{c}. Although Q-ball is stable against decay into heavy particles, such as squarks and sleptons, Q-ball may gradually decay into quarks and/or leptons from its surface. In this case, the final baryon number that contributes to the BBN is carried by quarks produced through the decay of Q-balls. However, in gauge mediated SUSY breaking models, a baryonic Q-ball with sufficiently large charge can be stable against decay into nuclei~(i.e.,~quarks)~\cite{dk}, while a leptonic Q-ball can decay into leptons since some leptons are much lighter than nuclei.

In our previous work~\cite{hyk}, we focused on gauge mediation type Q-balls that carry both baryon and lepton charges which can be formed after the Affleck-Dine baryogenesis with $u^cu^cd^ce^c$ flat direction, for instance. We found that it is possible for the Q-balls to be electrically charged, which we call charged or gauged Q-balls~\cite{gaugedqball}, due to the decay into charged leptons, until the electric charge becomes $\alpha^{-1}\simeq137$. Further decay is suppressed by Schwinger effect\footnote{This limit can also be understood by the discussion given by Feynman, where he pointed out that the ground state energy of the electron, which is the solution of the Dirac equation, becomes imaginary for $Z>\alpha^{-1}\simeq137$.}, and the charged Q-balls are stable by virtue of the stability of the baryonic component. 

We can expect that the charged Q-ball is dark matter in the present universe. The charged Q-balls capture the other charged particles and form atom-like, or ion-like objects which may make differences in the experimental signatures compared to neutral Q-balls. For instance, Q-balls can be detected by Super-Kamiokande~\cite{kensyutu, kttn} or IceCube~\cite{icecube}, which probe the KKST process~\cite{sipal} where the Q-ball absorbs quarks and emits pions of energy~$\sim1\mathrm{GeV}$. However, the charged Q-ball relics also experience the electromagnetic processes, so the detection of those processes are applicable. Various relevant experiments and upper bounds on the flux of the relics from the charged Q-balls are presented in Ref.~\cite{kensyutu}. The most stringent constraint comes from MICA experiment~\cite{mica}, where they claimed that the trail of heavy atom-like or ion-like object in $10^9$ years old ancient mica crystals is not observed. Its observation time is equal to the age of the mica, so that its constraint on Q-ball flux is much severer than those from other experiments. We found that the MICA constraint is more stringent than that from IceCube. 

In this paper, we assume that besides the positively charged Q-balls, nearly equal number of negatively charged anti-Q-balls of charge $|Z_Q|\sim\alpha^{-1}$ are formed as well, which is usually the case in gauge mediation model as we will mention in the next section. We make a rough predictions on the evolution of them, especially on when the recombination process with the other particle species in the universe takes place. Our main purpose is to identify the eventual relics in the present universe and apply the MICA constraint on the relics. Consequently, we find that the constraint translates into that on gauge mediation model parameter and reheating temperature.
\section{Q-balls in gauge mediated SUSY breaking model}
We consider the minimal gauge mediation model~\cite{mgm,mg2,mg3}, where SUSY is spontaneously broken by F-term of a field Z:
\begin{align}
\langle F_Z\rangle=F\neq0.
\end{align}
The soft breaking effect is mediated to the observable sector by messenger fields $\Psi$ and $\bar{\Psi}$, a pair of some representations and anti-representations of the minimal GUT group $SU(5)$, through the following interactions:
\begin{align}
W=kZ\bar{\Psi}\Psi+M_{\text{mess}}\bar{\Psi}\Psi,
\end{align}
where $k$ is a Yukawa constant, and $M_{\mathrm{mess}}$ is messenger mass.

An AD field in this model obtains the following potential~\cite{ks,mmgc,17} by the above soft breaking effect:
\begin{align}
V=V_{\text{gauge}}+V_{\text{grav}}=M_F^4\left(\log\left(\frac{\left|\Phi\right|^2}{M_{\text{mess}}^2}\right)\right)^2+m_{3/2}^2\left(1+K\log\left(\frac{\left|\Phi\right|^2}{M_{\ast}^2}\right)\right)\left|\Phi\right|^2+(\text{A-term}).
\end{align}
Here the first term comes from gauge mediation effect and
\begin{align}
M_F\simeq\frac{\sqrt{gkF}}{4\pi}
\label{eq:mfdef}
\end{align}
where $g$ generically means the gauge coupling of the standard model~\cite{mmgc}. 
The second term comes from gravity mediation effect and the gravitino mass $m_{3/2}$ is given by 
\begin{align}
m_{3/2}\equiv\frac{F}{\sqrt{3}M_{\text{P}}},
\end{align}
where $M_P=2.4\times10^{18}~\mathrm{GeV}$ is the reduced Planck mass. $K$ is a constant parameter that is related to beta function of mass of the AD field and is typically negative, satisfying $0.01\lesssim |K|\lesssim0.1$, and $M_*$ is the renormalization scale. The third term~(A-term)~is a CP an baryon number violating term which induces the AD rotation. 

A gauge mediation type Q-ball~\cite{21} is formed if $V_{\text{gauge}}$ dominates the potential after the AD field starts the oscillation. This is the case when
\begin{align}
\phi_{\text{osc}}<\phi_{\text{eq}}\simeq\sqrt{2}M_F^2/m_{3/2}
\label{eq:gad}
\end{align}
where $\phi\equiv\sqrt{2}|\Phi|$, and $\phi_{\text{osc}}$ denotes the field value at the beginning of the oscillation. We take $\phi_{osc}$ as a free parameter in this paper because it depends on an unknown higher-dimensional operater~\cite{hdo}. A typical charge of gauge mediation type Q-ball can be estimated by a linear approximation or numerical simulations, and it is given by~\cite{21}
\begin{align}
Q=\beta\left(\frac{\phi_{\text{osc}}}{M_F}\right)^4.
\label{eq:typicalcharge}
\end{align}
Here $\beta=6\times10^{-5}$ for an oblate orbit~($\epsilon\lesssim0.1$), where $\epsilon$ denotes the ellipticity of the field orbit. $\epsilon$ is usually small in the gauge mediation model, since the A-term, which induces the rotation, is proportional to $m_{3/2}$ which is typically small in the gauge mediation model. It is known that nearly the same number of Q-balls and anti-Q-balls are formed when the orbit is oblate, since the net baryon asymmetry is proportional to $\epsilon$ while the total energy is not suppressed by any powers of $\epsilon$~\cite{21}. In the following we focus on this case.

The mass, size and mass per unit charge of gauge mediation type Q-ball are given by
\begin{align}
M_Q&\simeq\frac{4\sqrt{2}\pi}{3}\zeta M_FQ^{3/4}\label{eq:mass}\\
R_Q&\simeq\frac1{\sqrt{2}}\zeta^{-1}M_F^{-1}Q^{1/4}\\
\frac{dM_Q}{dQ}&\simeq\pi R_Q^{-1},
\end{align}
respectively, where $\zeta$ is an $O(1)$ parameter~\cite{23,24}.

If we consider $u^cu^cd^ce^c$ direction, for example, only $e^c$ component can annihilate into $e^+$ inside Q-balls via gaugino and/or higgsino exchange interactions, for $dM_Q/dQ<m_p$, which is satisfied for a large enough Q-ball. Then electric charge is induced on the Q-ball, so we must consider the charged, or gauged Q-ball~\cite{gaugedqball}. From anti-Q-ball $e^-$ is emitted, so the electric charge is opposite to that of Q-ball. The electric charge can grow until $|Z_Q|\sim\alpha^{-1}$~\cite{hyk}, since the electric field at the surface of the Q-ball can only grow until $E\sim m_e^2/e$, above which $e^+e^-$ pair production occurs~(Schwinger limit). 
\section{Evolution of Charged Q-balls until present}
\label{sec:ev}
We assume that the positively charged Q-balls and negatively charged anti-Q-balls with $|Z_Q|\sim\alpha^{-1}$ are formed well before the BBN epoch, as shown in Ref.~\cite{hyk}. We consider the case that they eventually account for dark matter of the universe,~i.e:
\begin{align}
\frac{\rho_{\text{Q-ball}}}s+\frac{\rho_{\text{anti-Q-ball}}}s=\frac{\rho_{\text{DM}}}s\simeq4.4\times10^{-10}~\mathrm{GeV}
\label{eq:dm}
\end{align}
where $\rho_{\text{DM}}$ and $s$ are the dark matter energy density and entropy density in the present universe, respectively.

Charged Q-balls and anti-Q-balls are expected to capture other charged particles in the universe\footnote{\label{note1}Since the Q-balls are extremely heavy and the number densities are very small, we can safely ignore the (Q-ball)+(anti-Q-ball) reactions. For example, the typical gauge mediation type Q-ball, which is our case, has the baryon~(lepton)~charge $Q\sim10^{30}$ and mass $M_Q=M_FQ^{3/4}\sim 10^{28}\mathrm{GeV}$ for $M_F\sim10^6\mathrm{GeV}$.}. Since the number densities of (anti-) Q-balls must be extremely small in order to compose the dark matter, or to satisfy Eq.~(\ref{eq:dm}), due to their heavy masses~(See footnote~\ref{note1}), the recombination has almost no effect on the abundance of the other particle species, so that the BBN is not ruined. 

In Ref.~\cite{hyk}, we showed that for 
\begin{align}
Q\lesssim\left(\frac{4\sqrt{2}\pi\zeta M_F}{m_e}\right)^4\simeq2.5\times10^{39}\left(\frac\zeta{2.5}\right)^4\left(\frac{M_F}{10^6~\mathrm{GeV}}\right)^4,
\end{align} 
where $Q$ is baryon charge of the Q-ball, 
the size of the charged Q-ball with $Z_Q=\alpha^{-1}$ becomes smaller than the Bohr radius, so we approximate the electric potential in the outer region as the Coulomb type potential ($\sim1/r$). This allows us to treat the charged Q-balls as extremely heavy nuclei.

In the following, we roughly estimate when the recombination occurs for positively charged Q-balls and negatively charged anti-Q-balls, and discuss whether they become neutral~(like atoms) or charged~(like ions) eventually, so that we can identify the final relics in the present universe. We discuss positively charged Q-balls in the next subsection, and then discuss negatively charged anti-Q-balls in Sec.~\ref{sec:neg}. 
\subsection{Positively charged Q-balls}
The positively charged Q-balls can capture negatively charged particles, but they all annihilate except electrons at low temperature, so we consider the recombination with electrons\footnote{The discussion is somewhat similar to that in Ref.~\cite{champ}, where a singly charged dark matter is considered, but since we assume the charged dark matter with $|Q_{\text{electric}}|>100$, the results differ from that in Ref.~\cite{champ}.}. 

First, we estimate when the recombination starts, or 1S bound state is formed:
\begin{align}
Q+e^{-}\rightarrow 1S+\gamma,
\end{align}
where $Q$ means the charged Q-ball and $1S$ means that the charged Q-ball is surrounded by one electron.
The cross section from the free state to the 1S bound state, which we assumed as a hydrogen-type bound state, is evaluated in Ref.~\cite{atom}:
\begin{align}
\sigma v&=\frac{2^9\pi^2\alpha^{-1}}{3}\frac{E_{\text{bin}}}{m_e^3v}\left(\frac{E_{\text{bin}}}{E_{\text{bin}}+\frac12m_ev^2}\right)^2\frac{e^{-4\sqrt{\frac{2E_{\mathrm{bin}}}{m_ev^2}}\tan^{-1}\left(\sqrt{\frac{m_ev^2}{2E_{\text{bin}}}}\right)}}{1-e^{-2\pi\sqrt{\frac{2E_{\text{bin}}}{m_ev^2}}}}\\
&\simeq\frac{2^9\pi^2\alpha^{-1}}{3e^4}\frac{E_{\text{bin}}}{m_e^3v}
\end{align}
where $v$ is the relative velocity of the bare Q-ball and electron and $E_{\text{bin}}$ is binding energy. In the second line, we assumed that $m_ev^2/2\sim T\ll E_{\text{bin}}$, which corresponds to our case as we will see below. The thermal-averaged cross section is evaluated as~\cite{kohribbn}
\begin{align}
\langle\sigma v\rangle\simeq\frac{2^9\pi\alpha^{-1}\sqrt{2\pi}}{3e^4}\frac{E_{\text{bin}}}{m_e^2\sqrt{m_eT}}.
\end{align}
Then
\begin{align}
n_e\langle\sigma v\rangle/H\sim10^{12}\left(\frac{n_e/n_\gamma}{10^{-10}}\right)\left(\frac{3.2}{g_*}\right)^{1/2}\left(\frac{T}{\mathrm{GeV}}\right)^{1/2}
\label{eq:gamow}\end{align}
where $H$ is Hubble parameter and we assumed radiation dominated universe. This is larger than unity for a wide range of temperature, which allows us to use Saha's equation,
\begin{align}
\frac{n_{1S}n_{\gamma}}{n_{1S}^{EQ}n_{\gamma}^{EQ}}\sim\frac{n_{Q}n_e}{n_{Q}^{EQ}n_{e}^{EQ}},
\end{align}
where the subscripts represent number densities of corresponding particles and the superscript EQ means the thermal equilibrium value.
This leads to
\begin{align}
\frac{n_{1S}}{n_Q}\sim\frac{n_e}{n_\gamma}\left(\frac{T}{m_e}\right)^{3/2}\exp\left(\frac{E_{\mathrm{bin}}^{(1S,e)}}{T}\right).
\end{align}
Then, from the condition $n_{1S}/n_Q\sim1$, we can roughly estimate when 1S state is formed as
\begin{align}
T_c^{(1S,e)}\sim E_{\text{bin}}^{(1S,e)}/29.1\sim8.6~\mathrm{keV}
\end{align}
where we used $n_e/n_\gamma\sim O(10^{-10})$, and $E_{\text{bin}}^{(1S,e)}\simeq\alpha^2Z_Q^2Z_e^2m_e/2\simeq m_e/2$ since we assume the Coulomb type potential.

Next, we consider an era when the universe become cool enough so that $Q^{+1}$ ions can exist, which means that the Q-ball become completely neutral if it captures one more electron. Let us consider the following process:
\begin{align}
Q^{+1}+e^{-}\rightarrow Q^0+\gamma,
\label{eq:proc}
\end{align}
where $Q^0$ indicates the Q-ball neutralized by electrons. If we could use the Saha's equation, we would estimate when the final electron is captured, by the same analysis as above:
\begin{align}
T_c^{(\text{neu},e)}\sim E_{\text{bin}}^{(\text{neu},e)}/46.0\sim10^{-1}~\mathrm{eV}.
\end{align}
Here we assumed that the screening of the orbiting electrons make the binding energy smaller than that of hydrogen $13.6~\mathrm{eV}$, which is true for large enough~($Z>36$) elements~\cite{pdg}. We used a typical value $10~\text{eV}$.
Then we see that $T_c^{(\text{neu},e)}$ is slightly smaller than the proton-electron recombination temperature. This implies that electrons are recombined with protons and the number of free electrons rapidly decreases before the process of Eq.~(\ref{eq:proc}) occurred. Therefore, it is unlikely that the Q-ball ``atom" $Q^0$ forms at a temperature of $T=T_c^{(\text{neu},e)}$.

In addition, our assumption that $Q^{+1}$ ions can exist may not be correct in the first place, since even the temperature at which $Q^{+1}$ ions are formed may be smaller than the proton-electron recombination temperature, for instance if we naively assume that the second ionization energy is twice the first ionization energy. Thus we expect that $Q^{+O(1)}$ ions are likely to be formed, since we have no rigid information about ionization energies of $Z\sim137$ atoms.

\subsection{Negatively charged anti-Q-balls}
\label{sec:neg}
The negatively charged anti-Q-balls can capture the positively charged particles. Mainly protons and heliums remain after BBN, so we consider the recombination with them. 

Even if we take into account the difference in mass and charge of proton and helium compared to that of electron, $n_p\langle\sigma v\rangle/H$ and $n_{\text{He}}\langle\sigma v\rangle/H$ are still larger than unity. Therefore, we can use the Saha's equations:
\begin{align}
\frac{n_{1S}n_{\gamma}}{n_{1S}^{EQ}n_{\gamma}^{EQ}}\sim\frac{n_{Q}n_p}{n_{Q}^{EQ}n_{p}^{EQ}}
\label{eq:shp}
\end{align}
\begin{align}
\frac{n_{1S}n_{\gamma}}{n_{1S}^{EQ}n_{\gamma}^{EQ}}\sim\frac{n_{Q}n_{\text{He}}}{n_{Q}^{EQ}n_{\text{He}}^{EQ}}.
\end{align}
As before, we can roughly estimate when 1S state with proton is formed:
\begin{align}
T_c^{(1S,p)}\sim E_{\text{bin}}^{(1S,p)}/29.1\sim8.6~\mathrm{MeV},
\end{align}
where we used $E_{\text{bin}}^{(1S,p)}\simeq\alpha^2Z_{\bar{Q}}^2Z_p^2m_p/2\simeq m_p/2$ since, again, we are assuming the Coulomb type potential. Thus, proton is captured before helium is formed.

\begin{figure}[t]
  \includegraphics[width=\linewidth]{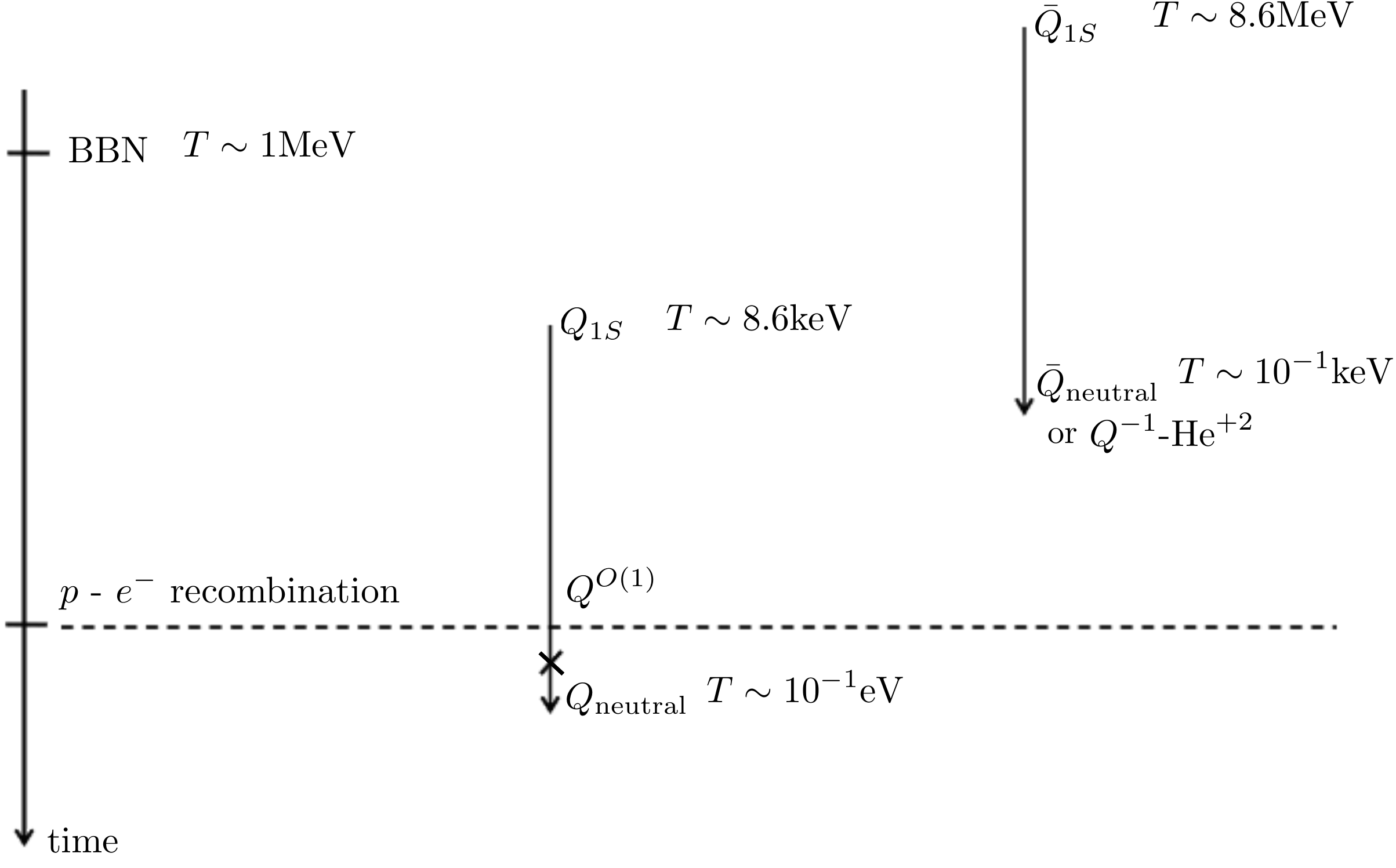}
 \caption{Evolution of positively (negatively) charged (anti-) Q-balls.}
\label{fig:time2}
\end{figure}

Next, we consider an era when the universe became cool enough so that $\bar{Q}^{-1}$ ions can exist, which means that the anti-Q-ball will become neutral if it captures one more proton. We can estimate when the final proton is captured:
\begin{align}
T_c^{(\text{neu},p)}\sim E_{\text{bin}}^{(\text{neu},p)}/46.0\sim10^{-1}~\mathrm{keV}.
\end{align}
Here, as the binding energy, we used $10~\mathrm{keV}$, which is the approximate value used in the previous subsection corrected by a factor of mass ratio $\sim m_p/m_e$. The anti-Q-balls become neutral much earlier than proton-electron recombination unlike the case of Q-ball-electron recombination, because the bound state becomes more difficult for photons to destruct, due to the heaviness of protons. There are yet plenty of free protons in this era, so $\bar{Q}^0$s are actually formed at $T_c^{(\text{neu},p)}$. 
Here we only considered the recombination with protons, but since the binding energy of helium to $\bar{Q}^{-1}$ is 16 times that of proton, $^4$He$^{+2}$ is easier to be captured. On the other hand, the protons start to be captured before the heliums are formed at $T\sim0.1~\text{MeV}$, so which element is mainly captured is non-trivial. 
If the helium is captured to $\bar{Q}^{-1}$, $\bar{Q}^{-1}$-He$^{+2}$ bound state may be formed as well, which is positively charged but cannot capture electrons for the same reason as $Q^{+1}$ in the previous section. We illustrate our results in Fig.~\ref{fig:time2}.

\section{Constraints from MICA experiment}
\label{sec:det}
Various experiments sensitive to electromagnetic processes and upper bounds on the flux of the charged Q-ball relics are presented in Ref.~\cite{kensyutu}. The most stringent comes from MICA experiment~\cite{mica}, 
\begin{align}
F\lesssim2.3\times10^{-20}~\mathrm{cm}^{-2}\text{s}^{-1}\text{sr}^{-1},
\end{align}
where they have not observed any trails of heavy atom-like or ion-like object in $10^9$ years old ancient mica crystals. Since detection time effectively becomes the age of the mica, its constraint is much severer than those from other experiments. The original purpose of the experiment is to detect the flux of magnetic monopoles which form the bound states with $^{27}$Al or other elements in the Earth, through a magnetic-dipole-magnetic-monopole interaction. These bound states can be regarded as supermassive atoms or ions, to which the detector is sensitive, since it is designed to be sensitive to atoms and ions with $Z\gtrsim10$. Relics from positively charged Q-balls can also be regarded as heavy $+O(1)$ charged ions with $Z\sim\alpha^{-1}=137$, and thus the detector is sensitive to them as well. Relics from negatively charged anti-Q-balls are like atoms or $+O(1)$ ions with about $10^3$ times heavier orbiting particles and with inverted charge sign between nucleus and orbiting particle. We do not know whether the detector is sensitive to these objects or not, since the stopping power in the experiment is fitted to the cases of realistic atoms and ions~\cite{stp}. But even if the detector is sensitive to the objects, the constraint on the mass will almost be the same as the case we neglected their existence, since $n_Q$ and $n_{\bar{Q}}$ are of the same order and the detector cannot identify the signals from different origins, only identifying that each stopping power has exceeded the threshold. Furthermore, the (anti-) Q-ball is so heavy that whether the orbiting particles are electrons or protons virtually has no effect on total mass. This situation is analogous to the case of monopoles, where we cannot identify whether $^{27}$Al or $^{55}$Mn is captured to the monopole from the experiment. In any case, dark matter flux is given by
\begin{align}
F&\simeq\frac{\rho_{\text{DM}\odot}}{M_Q}v
\end{align}
where $\rho_{\text{DM}\odot}$ is dark matter energy density near the solar system, and $v$ is the Virial velocity of the Q-balls.
Thus, using $\rho_{\text{DM}\odot}\sim0.3~\mathrm{GeV/cm^3}$ and $v\sim10^{-3}$, we obtain the following constraint on the mass of (anti-) Q-ball:
\begin{align}
M_Q\gtrsim3.9\times10^{26}~\mathrm{GeV}.
\label{eq:micac}
\end{align} 
This is quite a severe constraint since it almost reaches the typical mass of the Q-ball, and in turn constrains $\phi_{\text{osc}}$ via Eq.~(\ref{eq:typicalcharge}). In order to obtain the condition on $M_F$ and the reheating temperature $T_{\text{RH}}$, we derive a relation of $\phi_{\text{osc}}$ and reheating temperature $T_{\text{RH}}$ as
\begin{align}
\frac{\rho_{\text{DM}}}{s}\sim\frac{3T_{\text{RH}}}{4}\frac{M_Qn_\phi/Q}{3H_{\text{osc}}^2M_{\text{P}}^2}\sim\frac{3\pi}{2}\zeta\beta^{-1/4}T_{\text{RH}}\frac{\phi_{\text{osc}}^2}{M_{\text{P}}^2},
\label{eq:rdm}
\end{align}
where $n_\phi=m_{\text{eff}}\phi_{\text{osc}}^2$, $m_{\text{eff}}\simeq2\sqrt{2}M_F^2/\phi_{\text{osc}}$, $3H_{\text{osc}}\simeq m_{\text{eff}}$, and Eq.~(\ref{eq:mass}) are used~\cite{21}. $Q$ is the baryon charge of the Q-ball. 
Inserting the observational value $\rho_{\text{DM}}/s\simeq4.4\times10^{-10}~\mathrm{GeV}$~\cite{rhos}, we obtain
\begin{align}
\phi_{\text{osc}}\simeq4.3\times10^{12}~\mathrm{GeV}\left(\frac{\zeta}{2.5}\right)^{-1/2}\left(\frac{\beta}{6\times10^{-5}}\right)^{1/8}\left(\frac{T_{\text{RH}}}{\mathrm{GeV}}\right)^{-1/2}.
\label{eq:phitrh}
\end{align}
Using this relation and Eq.~(\ref{eq:typicalcharge}), Eq.~(\ref{eq:micac}) becomes
\begin{align}
T_{\text{RH}}\lesssim4.5\times10^{-2}~\mathrm{GeV}\left(\frac{\zeta}{2.5}\right)^{-1/3}\left(\frac{\beta}{6\times10^{-5}}\right)^{3/4}\left(\frac{M_F}{10^6~\mathrm{GeV}}\right)^{-4/3},
\label{eq:micac2}
\end{align}
which is indeed a constraint on $M_F$ and $T_{\text{RH}}$. 

\begin{figure}[!t]
  \includegraphics[width=\linewidth]{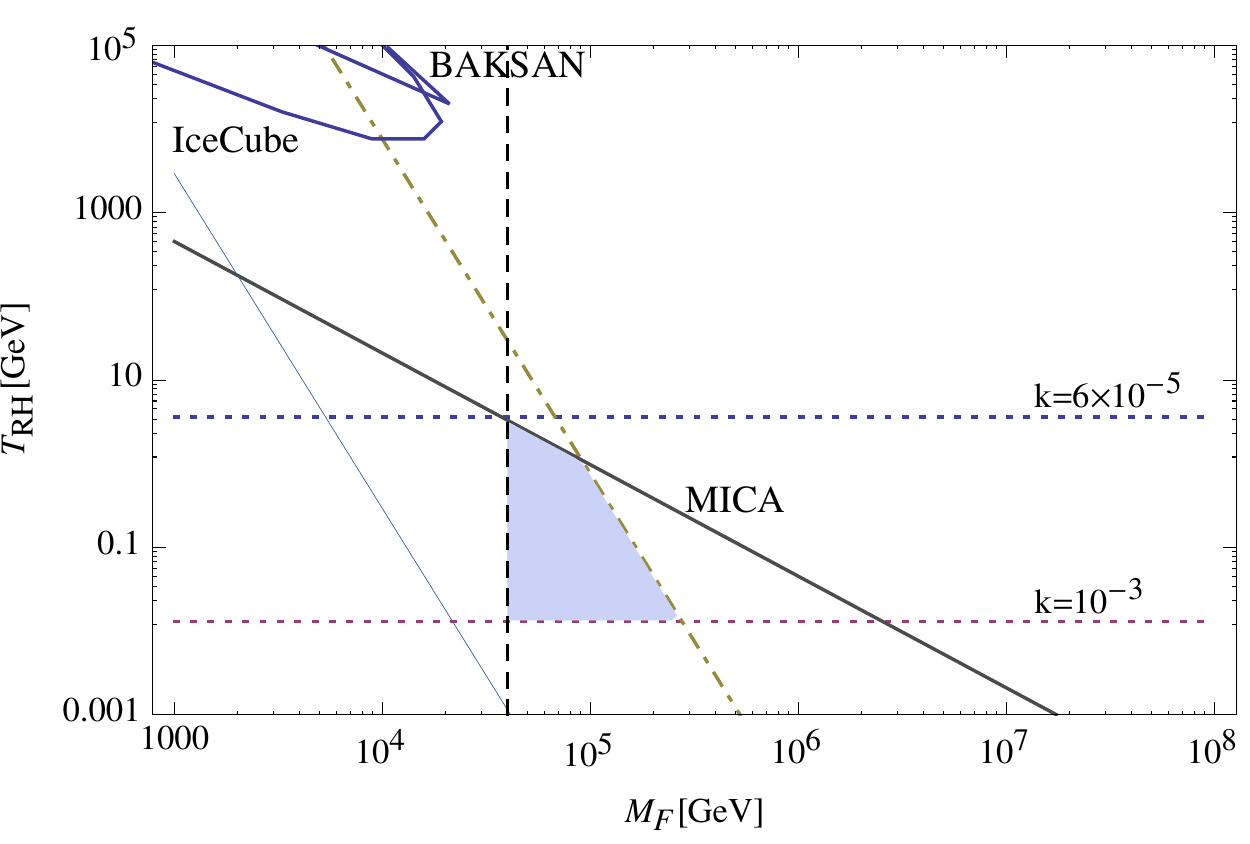}
 \caption{Allowed region for the gauge mediation type charged Q-ball as the dark matter~(shaded region), where we adopted $k=10^{-3}$. The solid line shows the upper bound Eq.~(\ref{eq:micac}), the dotted lines denote the lower bound Eq.~(\ref{eq:gad2}) for each value of $k$ shown, the dashed line corresponds to the $\Lambda_{\text{mess}}$-limit Eq.~(\ref{eq:h3}) with $g=1$, the dashed-dotted line is the upper bound Eq.~(\ref{eq:stc}), and the thin line represents the lower bound Eq.~(\ref{eq:bohr}). The IceCube and BAKSAN constraints are also represented at the upper left~\cite{icecube}.}
\label{fig:mftr}
\end{figure}

In Fig.~\ref{fig:mftr}, we illustrate the allowed region of $M_F$ and $T_{\text{RH}}$, using the analogous method to that in Ref.~\cite{icecube}, where the authors focused on IceCube and BAKSAN constraints on the neutral Q-ball dark matter.  
The MICA constraint Eq.~(\ref{eq:micac2}) corresponds to the solid line in the figure. 

The horizontal dotted lines indicate the condition that the gauge mediation effect dominates the potential of the Q-ball, which is given by Eq.~(\ref{eq:gad}).
Using Eq.~(\ref{eq:phitrh}), Eq.~(\ref{eq:gad}) becomes 
\begin{align}
T_{\text{RH}}\gtrsim1.3\times10^{-8}~\mathrm{GeV}g^{-2}k^{-2}\left(\frac{\zeta}{2.5}\right)^{-1}\left(\frac{\beta}{6\times10^{-5}}\right)^{1/4},
\label{eq:gad2}
\end{align}
which indeed corresponds to the horizontal dotted lines in Fig.~\ref{fig:mftr}. We adopted $k=10^{-3}$ as for identifying the allowed region. For $k\lesssim6\times10^{-5}$, there is no allowed region for gauge mediation type Q-ball, so we must consider the case of gravity mediation domination, where the horizontal lines become upper bounds. The Q-ball when the gravity mediation dominates is called new type Q-ball, which is beyond the scope of this paper.

It is known that the Higgs boson mass at around 126~GeV leads to \cite{mgm,ft}
\begin{align}
\Lambda_{\text{mess}}\equiv\frac{kF}{M_{\text{mess}}}\gtrsim5\times10^5~\mathrm{GeV}.
\label{eq:higgs}
\end{align}
Since the SUSY breaking scale is usually assumed to be small compared to the messenger mass
\begin{align}
kF<M_{\text{mess}}^2,
\end{align}
Eq.~(\ref{eq:higgs}) becomes
\begin{align}
\sqrt{kF}\gtrsim5\times10^5~\mathrm{GeV}.
\label{eq:h2}
\end{align}
Then, using Eq.~(\ref{eq:mfdef}), it reduces to 
\begin{align}
M_F\gtrsim4\times10^4g^{1/2}~\mathrm{GeV},
\label{eq:h3}
\end{align}
which corresponds to the vertical dashed line in Fig.~\ref{fig:mftr}.

For a Q-ball with baryon charge $bQ$, the electric charge $Q_{\text{electric}}$ must not be too large in order to be stable against baryonic decay~\cite{hyk}, which leads to 
\begin{align}
bm_p&>\frac{dE}{dQ}\label{eq:dedbph}\\
&\simeq \omega_0+\frac{e^2Q_{\text{electric}}}{4\pi R_0}\nonumber\\
&=\omega_{0}+\frac{e^2Q_{\text{electric}}}{4\pi^2}\omega_{0}\nonumber\\
&=\left(\sqrt{2}\pi+\frac{e^2Q_{\text{electric}}}{2\sqrt{2}\pi}\right)\zeta M_FQ^{-1/4}.
\end{align}
Thus, we obtain
\begin{align}
Q_{\text{electric}}\lesssim\frac1{\alpha}\left(bm_p\frac1{\sqrt{2}}\zeta^{-1}M_F^{-1}Q^{1/4}-\pi\right).
\end{align}
where we used thin-wall approximation on charged Q-ball~\cite{gaugedqball} in the second line, and $b=1/3$ for the $u^cu^cd^ce^c$ flat direction, for example.
The condition that above is satisfied for the eventual electric charge $Z_Q\sim\alpha^{-1}$ becomes 
\begin{align}
Q\gtrsim\left(\frac{\sqrt{2}(\pi+1)\zeta M_F}{bm_p}\right)^4\simeq
3.7\times10^{30}\left(\frac{\zeta}{2.5}\right)^4\left(\frac{b}{1/3}\right)^{-4}\left(\frac{M_F}{10^6~\mathrm{GeV}}\right)^4.
\end{align}
Using Eq.~(\ref{eq:typicalcharge}) and Eq.~(\ref{eq:rdm}), we obtain
\begin{align}
T_{\text{RH}}\lesssim7.5\times10^{-5}~\mathrm{GeV}\left(\frac{\zeta}{2.5}\right)^{-3}\left(\frac{\beta}{6\times10^{-5}}\right)^{3/4}\left(\frac{b}{1/3}\right)^2\left(\frac{M_F}{10^6~\mathrm{GeV}}\right)^{-4},
\label{eq:stc}
\end{align}
which corresponds to the dashed-dotted line in Fig.~\ref{fig:mftr}.

Finally, as mentioned in the previous section, we consider the (anti-) Q-ball smaller than Bohr radius so that the potential the external particles experience can be approximated into Coulomb-type potential. Thus, the following condition must be satisfied,
\begin{align}
Q\lesssim\left(\frac{4\sqrt{2}\pi\zeta M_F}{m_e}\right)^4\simeq2.5\times10^{39}\left(\frac\zeta{2.5}\right)^4\left(\frac{M_F}{10^6~\mathrm{GeV}}\right)^4
\end{align}  
which becomes 
\begin{align}
T_{\text{RH}}\gtrsim2.9\times10^{-9}~\mathrm{GeV}\left(\frac{\zeta}{2.5}\right)^{-3}\left(\frac{\beta}{6\times10^{-5}}\right)^{3/4}\left(\frac{M_F}{10^6~\mathrm{GeV}}\right)^{-4},
\label{eq:bohr}
\end{align} 
where we used, again, Eqs.~(\ref{eq:typicalcharge}) and~(\ref{eq:rdm}). This corresponds to the thin line in Fig.~\ref{fig:mftr}. We see that this condition is automatically satisfied by virtue of the other conditions, which means that the other conditions constrain the (anti-) Q-balls to be smaller than Bohr radius so that the Coulomb type potential can be used.

The IceCube and BAKSAN constraints are shown at the upper left for comparison. We see that the MICA constraint is more stringent than that from IceCube and BAKSAN, where only the KKST process is probed. This makes the allowed region smaller.

\section{Conclusions and discussion}
\label{sec:conc}
In this paper, we considered nearly equal number of gauge mediation type charged Q-balls and anti-Q-balls with charge of $\alpha^{-1}$, as expected from Ref.~\cite{hyk}, and discussed how they evolve in time and what kinds of objects become relics in the present universe. We roughly estimated that the positively charged Q-balls start to capture electrons at $T\sim8.6~\mathrm{keV}$, and eventually become $Q^{+O(1)}$ ions, since the number of free electrons decreases due to the recombination with protons before the Q-ball can capture further electrons. The negatively charged (anti-) Q-balls start to capture protons at $T\sim8.6~\mathrm{MeV}$, and after the BBN, helium and the other light elements can also be captured. Anti-Q-balls are neutralized by the final proton or positively charged by the final $^4$He at $T\sim10^{-1}~\mathrm{keV}$, since it is earlier than the proton~(helium) - electron recombination so the plenty of protons and heliums exist.

The $Q^{+O(1)}$ ions from the Q-balls can be treated as ordinary ions since the potential is nearly Coulomb type. Thus, they are detectable by MICA experiment, where no trail of heavy atom-like or ion-like object is observed in $10^9$ years old ancient mica crystals. This gives the most stringent constraint on the flux of the objects, which are assumed to compose the dark matter, among the constraints obtained in Ref.~\cite{kensyutu}, since the detection time is extremely long which is essentially the age of the mica. We translated the constraint into that on the gauge mediation model parameter $M_F$ and reheating temperature $T_{\text{RH}}$, as done in Ref.~\cite{icecube} for the IceCube and BAKSAN constraints on the neutral Q-ball dark matter. As a result, we found that the MICA constraint is more severe than that from IceCube and BAKSAN, so the allowed region in $M_F$ - $T_{\text{RH}}$ becomes smaller.

Identifying the relics from Q-balls and those from anti-Q-balls observationally, which includes examining the properties of (anti-Q-ball)+(nucleus) bound state, will be one of our future tasks. Also, we assumed $u^cu^cd^ce^c$-like flat direction which includes only electron as the leptonic component. However, if we consider the direction which includes neutrino component, for example $QQQL$, the scenario may differ from what we pursued so far, and interesting in its own right. For instance, due to the decay into neutrinos, the Q-ball may have $SU(2)$ charge, and we may need a fundamental theory of non-abelian gauged Q-ball. Finally, the Q-ball when the gravity mediation dominates is not discussed in this paper, which is called new type Q-ball. New type Q-ball has different properties from gauge mediation type Q-ball, including its mass, size, and typical charge etc., and identifying the relics and investigating the possibility as dark matter will also be of our future interest.


\vspace{1cm}

\section*{Acknowledgments}
This work is supported by MEXT KAKENHI Grand Number 15H05889~(M.K.) and JSPS KAKENHI Grand Number 25400248~(M.K.), 
World Premier International Research Center Initiative
(WPI Initiative), MEXT, Japan,
and the Program for the Leading Graduate Schools, MEXT, Japan (M.Y.).
M.Y. acknowledges the support by JSPS Research Fellowships for Young Scientists, No.25.8715.

\vspace{1cm}



\end{document}